\documentclass[11pt,a4paper,twocolumn]{article}

\usepackage{flushend}
\usepackage{lineno}
\usepackage{comment}
\usepackage{authblk}
\usepackage{graphicx}
\usepackage[explicit]{titlesec}
\usepackage[labelfont=bf,labelsep=endash,font=footnotesize]{caption}
\usepackage{tabu}
\usepackage{xcolor}
\usepackage[
    backend=biber, 
    natbib=true,
    style=numeric,
    maxnames=50,
    sorting=none
]{biblatex}
\usepackage[hang]{footmisc}
\usepackage[normalem]{ulem}
\usepackage[top=2.5cm, bottom=2.8cm, left=1.5cm, right=1.5cm]{geometry}
\usepackage{abstract}
\usepackage{mathtools}
\usepackage{balance}
\usepackage{wrapfig}
\usepackage{multirow}

\makeatletter
\renewcommand\AB@affilsepx{, \protect\Affilfont}
\makeatother

\providecommand{\keywords}[1]{\textbf{Keywords}\ \ \textendash\ \   #1}

\renewcommand{\figurename}{Fig.}

\titleformat{\section}{\large\bfseries}{\thesection.}{1em}{\MakeUppercase{#1}}
\titlespacing*{\section}{0pt}{12pt}{6pt}

\titleformat{\subsection}{\large}{\thesubsection}{1em}{#1}
\titlespacing*{\subsection}{0pt}{12pt}{6pt}

\titleformat{\subsubsection}{\large\itshape}{\thesubsubsection}{1em}{#1}
\titlespacing*{\subsubsection}{0pt}{12pt}{6pt}

\newcommand{\ITUurl}[1]{\textcolor{blue}{\urlstyle{same}\url{#1}}}

\newcommand{\ITUpar}{\vspace{8pt}\par}

\renewenvironment{abstract}
               {\list{}{
               \setlength{\rightmargin}{0mm}
               \setlength{\leftmargin}{0mm}
               \vspace{-0.25in}
                \item[\textit{\textbf{\hspace{22pt}Abstract  }}  \textendash]\relax}}
               {\endlist}

\def\starttable{\vspace{6pt}\begin{table}[ht]\center}

\def\startfigure{\vspace{6pt}\begin{figure}[ht]\center} 

\renewcommand\theequation{\textit{(\arabic{equation})}}
\makeatletter
\def\tagform@#1{\maketag@@@{\ignorespaces#1\unskip\@@italiccorr}}
\makeatother

\title{\large{\textbf{\uppercase{Networking for the Metaverse: The Standardization Landscape}}}}

\author{\normalsize{Cedric Westphal}}
\affil{\normalsize{Futurewei Technologies Inc.\\2220 Central Expressway, Santa Clara, CA, USA}} 

\author{\normalsize{Jungha Hong and Shin-Gak Kang}}
\affil{\normalsize{Electronics and Telecommunications Research Institute\\218 Gajeong-ro, Yuseong-gu, Daejeon, Korea}}

\author{\normalsize{Leonardo Chiariglione}}
\affil{\normalsize{MPAI\\5 Cours des Bastions,
CH-1205, Geneva, Switzerland}}

\author{\normalsize{Tianji Jiang}}
\affil{\normalsize{China Mobile Tech., 1525 McCarthy Blvd., Milpitas, CA 95035, USA}}

\date{\vspace{-12pt}{\small NOTE: Corresponding author: cedric.westphal@futurewei.com} \\  \endgraf\rule{\textwidth}{1pt}}

\begin{document}

\twocolumn[

\begin{@twocolumnfalse}
\maketitle
\thispagestyle{empty}

\begin{abstract}
\textit{New applications are being supported by current and future networks. In particular, it is expected that Metaverse applications will be deployed in the near future, as 5G and 6G network provide sufficient bandwidth and sufficiently low latency to provide a satisfying end-user experience. However, networks still need to evolve to better support this type of application. We present here a basic taxonomy of the metaverse, which allows to identify some of the networking requirements for such an application; we also provide an overview of the current state of balthe standardization efforts in different standardization organizations, including ITU-T, 3GPP, IETF and MPAI.}
\end{abstract}

\ITUpar
\keywords{Metaverse, AR/VR, end-to-end latency guarantee, future Internet, high precision networking, ITU-T, 3GPP, IETF, MPAI }

\ITUpar 
\ITUpar

\end{@twocolumnfalse}
]


\section{Introduction} 
\label{sec:intro}

New applications are being supported by current and future networks. In particular, it is expected that Metaverse applications will be deployed in the near future, as 5G and 6G network provide sufficient bandwidth and sufficiently low latency to provide a satisfying end-user experience. However, networks still need to evolve to better support this type of application.\ITUpar

Figure~\ref{fig:LBQ} (from~\cite{GSMA-5G-short}) describes the latency and bandwidth requirements for a number of emerging applications. The rectangular area shaded in gray holds the services that can be deployed on a 5G network. Augmented reality (AR), virtual reality (VR) stand at the edge of this domain, with both latency and bandwidth requirements that are at the limit of 5G networks support. \ITUpar

The Metaverse applications would provide some combination of AR, VR, social media, video-conferencing and wold inherit the latency and bandwidth requirements from these existing applications. From this, we can derive a few observations. \ITUpar

The first observation is that, of course, the requirements will vary based upon how the Metaverse is instantiated. For instance, legless cartoonish avatars may not require as much bandwidth as a lifelike rendition; haptic interfaces may require much stringent latency and jitter than a purely visual interface. As such, the requirements of the Metaverse upon the network will vary a lot upon what the Metaverse is, and what it looks and feels like. \ITUpar

The second observation is that, if we assume similar requirements than AR/VR for instance, then a Metaverse application will find itself at the edge of what 5G networks can support. Indeed, there are no successful commercial deployments as of this writing that run over 5G networks. This means that for a Metaverse application to work, it will need support from the network, either in the form of some edge node support, some distributed implementation, some enhancement to the network to provide the proper level of QoE, etc.\ITUpar

Both observations underline the need for standardization, in particular at the network layer. Commonly agreed definitions, use cases, or requirements are a prerequisite for standardization. Further, since the involvement of the network is necessary, standardization will be required at this layer as well. \ITUpar

In this paper, we attempt to provide an overview of the standardization landscape for Metaverse applications. We first attempt to specify this landscape by considering the different potential instances for a Metaverse. We then discuss ongoing efforts in different Standards Development Organizations (SDOs), including ITU-T, 3GPP, IETF, etc. \ITUpar

\startfigure
\includegraphics[width=\columnwidth]{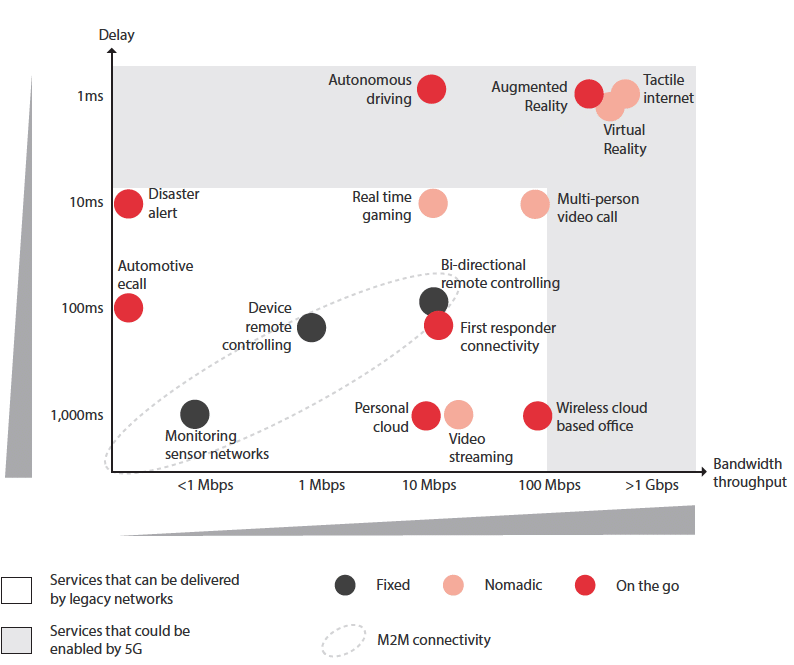}
\caption{Latency/Bandwidth Requirements for Emerging Applications in 5G Networks~\cite{GSMA-5G-short}}
\label{fig:LBQ} 
\end{figure}

This paper is organized as follows:\ITUpar
\begin{itemize}
    \item Section~\ref{sec:related} goes over the existing literature in the space of Metaverse requirements for networking. 
    \item Section~\ref{sec:taxonomy} presents a taxonomy of the different types of Metaverse applications, and attempts a definition of the Metaverse. From this taxonomy, we infer some of the requirements to support a Metaverse application.
    \item Section~\ref{sec:ITU} goes over Metaverse standardization in the ITU-T.
    \item Section~\ref{sec:3GPP} considers similar efforts in 3GPP.
    \item Section~\ref{sec:IETF} looks at IETF efforts towards Metaverse.
    \item Section~\ref{sec:MPAI} considers similar efforts in MAPI.
    \item Section~\ref{sec:IEEE} considers other SDOs involved in standardizing the Metaverse.
    \item Section~\ref{sec:table} compiles a table summary of the different efforts.
    \item Finally, Section~\ref{sec:conclusion} closes the paper with some concluding remarks.
\end{itemize}

While our overview is not exhaustive, as networking research has a long history and covers many domains, we hope to point out some important directions for future investigation, future protocol design, and maybe, future deployment in an evolved Internet.

\section{Related Works}
\label{sec:related}

The term Metaverse famously originated in the Neal Stephenson best seller, Snow Crash~\cite{stephenson1992snow}. It has been prominent in popular culture, as exemplified by the success of Ready Player, One~\cite{cline2011ready} and technological advances\cite{eckert2022overview} are closer and closer to bringing these concepts to reality (while being cautious of the potential downsides of such an application~\cite{chohan2022metaverse}). Sustainability considerations may motivate and drive uses of Metaverse applications as a substitute for energy-costly in-person collaboration\cite{clemm2022challenges,irtf-nmrg-green-ps-01}.\ITUpar

The Metaverse will depend on new interfaces being deployed. All-sense (and beyond) human/machine and human/human interfaces are one such key area of upcoming technology development. Holographic communications include holographic environment capturing and remote holographic rendering. It is the next stage of visual communications beyond AR and VR and it requires network throughput that is orders of magnitude higher than today's 2D/3D and AR/VR visual communications~\cite{westphal2017challenges,he2018network-short,he2018joint}, as well as better guaranteed performance for the quality of holographic rendering required to achieve the desired benefits, lifelike visual remote presence.\ITUpar

A broad survey of the Metaverse~\cite{wang2023survey-short} encompasses the current development status in multiple countries, the technological and social challenges. Related to our work is that it identifies multi-technology convergence as a key challenge, relating to the five main technological issues of the Metaverse: 1) communication and computing infrastructure, 2) management technology; 3) fundamental common technology (AI, spatio-temporal consistency, security \& privacy); 4) VR object connection; and 5) VR space convergence.\ITUpar

While the underlying technologies could be mapped to different quadrants, there is no debating that the integration of these multiple technology is a challenge that calls for inter-operability across multiple domains. \ITUpar

\cite{khan2023metaverse-short} comprehensively surveys the Metaverse environment from a wireless perspective; indeed the standardization discussed in that paper is focused on ML-enabled
wireless metaverse, which is a much narrower scope than our work. \ITUpar

\cite{hyun2023study-short} is closely related to our work in its attempt to describe the standardization landscape. However, this brief note does not cover the same SDOs and not with the same level of detail.

\cite{yang2023recommendations} considers standardization from the point of view of governance. Indeed, while the impact of Metaverse cannot be fully known before the technology is deployed, a governance structure is nonetheless required. It finds that effective metaverse governance should comprise three elements: 1) a formulation of technical standards; 2) compatibility of these standards; 3) they must be secure. The only SDO mentioned in the paper is that of  IEEE P2048 Standard for Metaverse: Terminology, Definitions, and Taxonomy, originated in the IEEE Metaverse Standards Committee (CTS/MSC). In a similar vein, \cite{rawat2023metaverse} mentions also Metaverse standardization by focusing on IEEE. It also mentions ISO/IEC 23005. Our work casts a wider net to include other standardization bodies.

\section{Taxonomy and Requirements}
\label{sec:taxonomy} 

First we need to define what we mean by “metaverse” and introduce some taxonomy to help us specify what can or cannot be standardized. \ITUpar

\subsection{Definition}
\label{sec:def}

We present here four possible definitions. We do not settle on one specific definition, as it is not our scope to offer a definitive definition of the metaverse, or to settle any debate about what is/what isn't a metaverse. Rather we see in the different definitions a different set of implications for the design of the application. \ITUpar

{\bf Definition 1:} (Damar~\cite{damar2021metaverse}) “a 3D virtual shared world where all activities can be carried out with the help of augmented and virtual reality services.”\ITUpar

{\bf Definition 2:} [Meta, 2022] “an integrated immersive ecosystem where the barriers between the virtual and real worlds are seamless to users, allowing the use of avatars and holograms to work, interact and socialize with simulated shared experience.”\ITUpar

{\bf Definition 3:} (John Riccitiello~\cite{riccitiello-def}) “the next generation Internet that is always real-time and mostly 3D, mostly interactive, mostly social and mostly persistent.” \ITUpar

Note that the first definition is an extension of an AR/VR framework; the second definition includes an ecosystem, which assumes a set of API to integrate multiple elements into the ecosystem; the last definition views the metaverse as the replacement of the Internet, that is a global scale application that supports an unlimited range of applications and functions, with a requirement of persistence. \ITUpar

As we will discuss the ITU-T effort below, we note that ITU-T has provided its own preliminary definition for further discussion and harmonization with other SDOs:

{\bf Definition 4:} (ITU-T~\cite{ITU-def}) “The Metaverse is an integrative and unified ecosystem of virtual worlds, which is based on inter-operable Internet-based and enhanced reality systems, and offers immersive experience to individuals during their digital and synchronize interactions, and new value generation opportunities to organizations.”

\subsection{Taxonomy}
\label{sec:tax}

As with the definition of the metaverse, we can try to better define what a metaverse is by way of a taxonomy that differentiates according to different criteria. The dimensions that we consider are listed below. This list is inspired by~\cite{dwivedi2022metaverse-short} but includes additional dimension.\ITUpar

{\bf Environment:} the environment can be realistic, unrealistic, fused; the more realistic (or detailed) the more bandwidth is required; conversely, some unrealistic environment can be generated and rendered from some basic models that can be distributed ahead of time. The environment can also be generated anew every time, or have some permanence. In the latter case, it can be cached at the edge or on the device.\ITUpar

{\bf Interface:} the environment can be interacted with through an interface that ranges from a simple phone screen to a 3D head-mounted display (HMD), from a window into the virtual world into an immersive experience; other physical methods to interface the virtual environment (such as haptics) can be included as well.\ITUpar

{\bf Interaction:} the level of interaction can be specific to the virtual environment. It can be in one extreme a solitary experience (such as playing game against a computer) and extend to social networking, and/or work collaboration. The granularity of the interaction also impacts the infrastructure requirements.\ITUpar

{\bf Security:} it is paramount to protect the security and privacy of the experience. This includes data security, privacy, software/hardware/network security. Further, the granularity of the security may include several layers, as for instance, only a given set of participants can access a given shared metaverse; and within this metaverse, only a subset can have access to objects or rooms within.\ITUpar

{\bf Centralization:} this is not a characteristic of the metaverse itself, rather a design choice on how to deploy such an application over some infrastructure. However, this choice has an impact on the infrastructure and needs to be considered. Centralization of the metaverse, by hosting it on a specific set of servers and have clients connect to these servers, facilitates some aspects of the metaverse: for instance, it requires N connections, where N is the number of users; it facilitate access controls, as per the "security" item above.\ITUpar

A fully distributed architecture that is fully meshed would require $N^2$
(potentially multi-cast) connections; further, these connections would need to be time-synchronized. However, the latency of a direct path would always be faster than a triangular routing through a server, and therefore the interactions would be quicker. This list of dimension is not necessary exhaustive, as other dimensions may arise in the deployment of the Metaverse. \ITUpar

Note that decentralization is sometimes viewed as a mechanism to bring distributed ledgers/blockchains into the metaverse. This is orthogonal to the scope of the paper that focuses on the networking layer mostly. 

\subsection{Levels of Interoperability}
\label{sec:inter}

\startfigure
\includegraphics[width=\columnwidth]{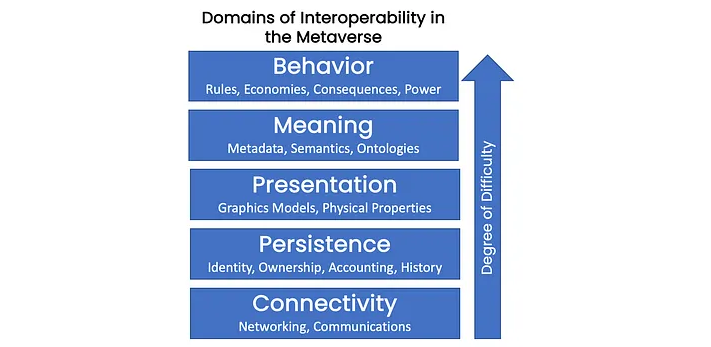}
\caption{Interoperability Layers (from~\cite{radoff-interoperability})}
\label{fig:Inter} 
\end{figure}

Jon Radoff~\cite{radoff-interoperability} wrote about the different layers for interoperability. This is captured in Figure~\ref{fig:Inter} from~\cite{radoff-interoperability}. 

While all layers are required to provide a successful Metaverse experience, most of the standardization work we describe is at the connectivity layer. 

We will briefly describe other standardization efforts at other interoperability layers. 

\section{ITU-T}
\label{sec:ITU}

The International Telecommunication Union (ITU) has been at the forefront of global technology standardization for over 150 years. As shown in Figure~\ref{fig:ITU}, the ITU has 193 member states and more than 900 private sector entities, universities, and international and regional organizations as members, serving as the primary international platform for all stakeholders to build consensus on the important and pressing ICT technical and regulatory issues facing our society today. 

\startfigure
\includegraphics[width=\columnwidth]{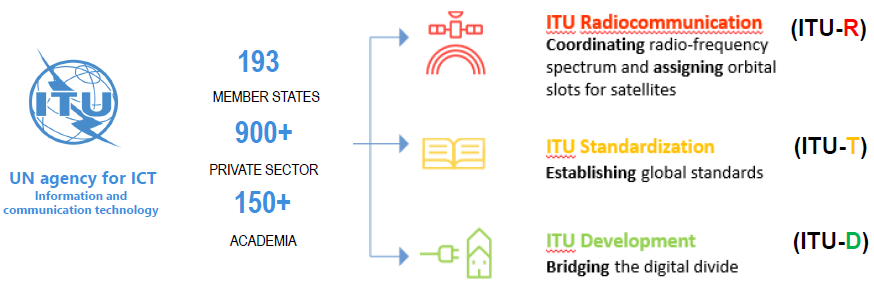}
\caption{ITU structure}
\label{fig:ITU} 
\end{figure}

Metaverse standardization discussions had initiated in several ITU-T Study Groups (SGs), including SG16, SG17, and SG20, and other SGs had also shown interest in metaverse. In particular, ITU-T SG16 (Multimedia and related digital technologies) established a correspondence group on metaverse (CG-Metaverse) in January 2022 to discuss metaverse standardization issues. The CG-Metaverse also discussed the need to establish an ITU-T focus group on metaverse (FG-MV). Finally, the ITU-T FG-MV was established under the ITU-T Telecommunication Standardization Advisory Group (TSAG) in December 2022, as the scope of metaverse standardization is not limited to a specific working group~\cite{bueti2023executive}.

The objective of FG-MV is to support pre-standardization activities, including the following:~\cite{ITU-T-FG-MV-ToR}

\begin{itemize}
    \item To study terminology, concepts, vision and ecosystem.
    \item To identify and study the enabling technologies, their evolution and key tasks for standardization purposes, including multimedia, network optimization, connectivity, interoperability of services and applications, security, protection of personally identifiable information, quality (including bandwidth), digital assets (e.g., digital currencies), IoT, accessibility, digital twin and environmental sustainability.
    \item To study and gather information to develop a pre-standardization roadmap.
    \item To build a community of experts and practitioners to unify the concepts, develop common understandings, so that it be benefiting not only the ITU standardization scene but also the global community.
    \item To identify stakeholders with whom ITU-T could collaborate and establish liaisons and relationships with other organizations that could contribute to the pre-standardization activities and identify potential collective action and specific next steps.
    \item To stimulate international collaboration, to share knowledge and best practices, and to explore the opportunities and challenges related to interoperability.
    \item To provide a platform to share findings and for dialogue on economic, policy and regulatory implications of metaverse related to telecommunication/ICT.
\end{itemize}

We note that the ITU-T focus group is an instrument to provide an additional working environment for the quick development of standards in specific areas and it is acting as an incubator for a future emerging issue. The focus group opens to all experts with an interest and even to non ITU members. The documents developed in the focus groups are called deliverables and are categorized as Technical Specifications (TSs) or Technical Reports (TRs) depending on their nature.  Deliverables are later submitted to the ITU-T study groups for further review and adoption as Recommendations, Supplements or Technical reports.

The FG-MV was originally scheduled to operate for one year, but a proposal to extend it for another year will be discussed at the last FG-MV meeting in December 2023. This will be decided at the next TSAG meeting in early 2024. Oversight and reporting of the FG-MV's activities will be directed to the ITU Telecommunication Standardization Advisory Group~\cite{ITU-T-FG-MV}.

The FG-MV is organized into Working Groups (WGs) and Task Groups (TGs). WGs are responsible for standardization work in a major area and TGs conduct discussions and develop one or more documents focusing on more specific topics that fall within the scope of the WG's work. There are currently 9 WGs and 19 TGs approved as shown in Table~\ref{FG-MV structure}, but the structure may change in the future as standardization work is progressed. The details of WGs and TGs can be found at~\cite{ITU-T-FG-MV-WP}.
 
\begin{table*}
\centering
\caption{Structure of ITU-T FG-MV}
\label{FG-MV structure}
\begin{tabular}{ |p{4cm}|p{12cm}| } 
\hline
 Working Groups & Task Groups \\
\hline
\hline
\multirow{3}{8em}{WG 1 - General} & - Terminology \& definitions \\ 
& - Pre-standardization for the CitiVerse  \\ 
& - Implications for people in the metaverse  \\ 
\hline
\hline
\multirow{7}{8em}{WG 2 - Applications \& Services} & - Media coding \\ 
& - Generative Artificial Intelligence in the metaverse \\
& - Embodied Artificial Intelligence for metaverse \\
& - Metaverse tourism \\
& -	Medical metaverse \\
& - Power metaverse \\
& -	Industrial metaverse \\
\hline
\hline
 \multicolumn{2}{|l|}{WG 3 - Architecture \& Infrastructure} \\
\hline
\hline
\multicolumn{2}{|l|}{WG 4 - Virtual/Real World Integration} \\
\hline
\hline
\multicolumn{2}{|l|}{WG 5 - Interoperability} \\
\hline
\hline
\multirow{4}{8em}{WG 6 - Security, Data \& PII Protection} & - Cybersecurity \\
& - Building confidence and security in the metaverse \\
& - Child online protection \\
& - Issues on trustworthiness related to the metaverse \\
\hline
\hline
\multicolumn{2}{|l|}{WG 7 - Economic, regulatory \& competition aspects} \\
\hline
\hline
\multirow{4}{8em}{WG 8 - Sustainability, Accessibility \& Inclusion} & - 	Sustainability \\
& - Accessibility \& inclusion \\
& -	Design criteria and metrics with incentives for sustainable metaverse \\
& -	Metaverse social safety \\
\hline
\hline
WG 9 - Collaboration & - Gap analysis \\
\hline
\end{tabular}
\end{table*}

Since the FG-MV has established, a total of 59 standardization work items have been adopted to develop deliverables, and the first official deliverable, ‘Exploring the metaverse: opportunities and challenges’, was approved at the second meeting of FG-MV in July 2023. The rapid publication of an official report is highly unusual and demonstrates the rapid pace of FG-MV standardization work. The deliverable explores the background of the metaverse, including its history, ecosystem and development stages, challenges and opportunities. Figure~\ref{fig:Challenges} shows the opportunities and challenges encountered in the metaverse, from interoperability and digital identity to sustainability and regulations~\cite{ITU-T-FG-MV-TR}.

\startfigure
\includegraphics[width=\columnwidth]{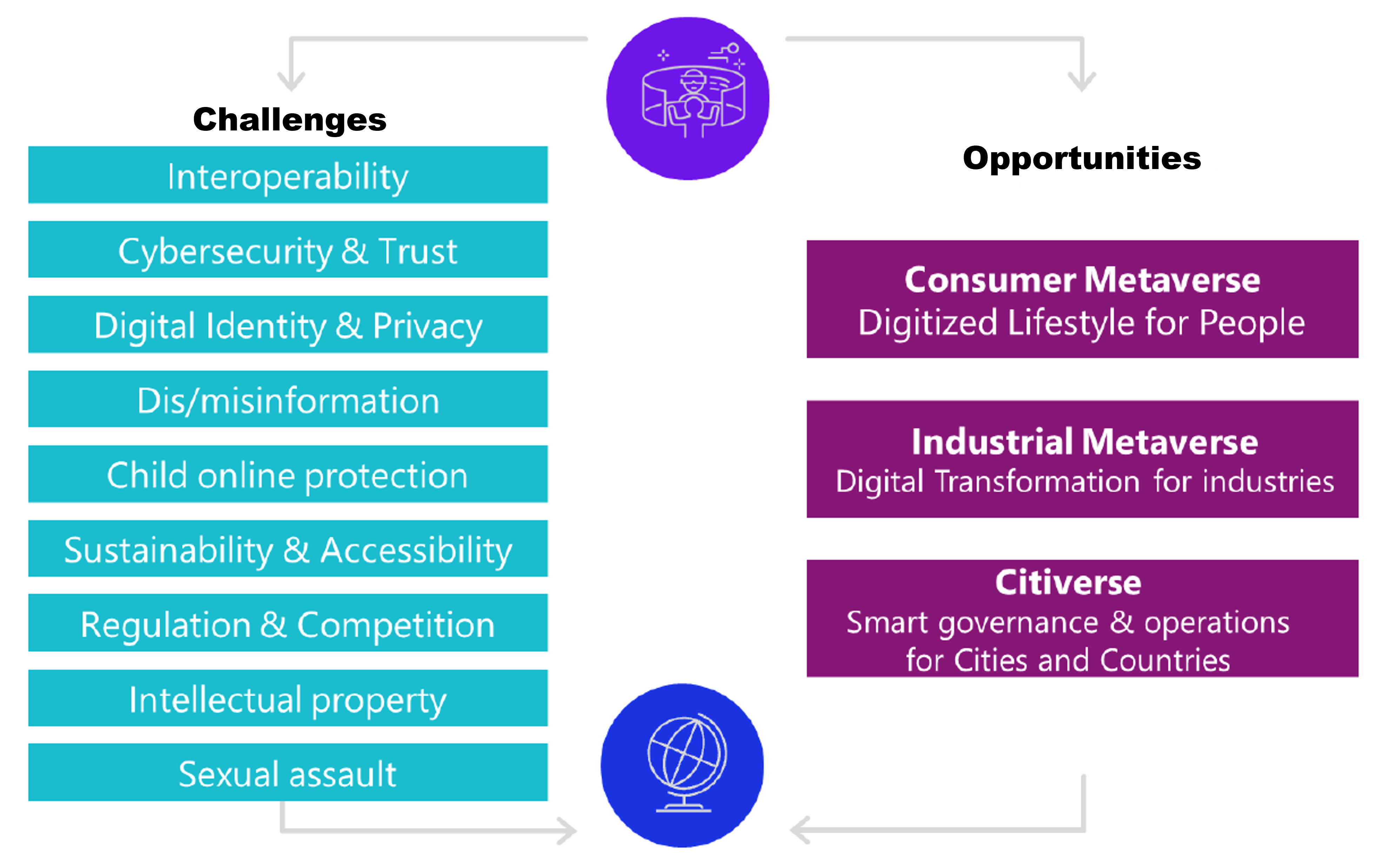}
\caption{Opportunities and Challenges of metaverse}
\label{fig:Challenges} 
\end{figure}

ITU-T FG-MV meetings are scheduled for March, July, October, and December 2023. As of September 2023, the first and second meetings have been held. Over 650 participants attended the first meeting in Riyadh, Saudi Arabia and over 2,000 attended the second meeting in Shanghai, China, both on-site and online, which is unprecedented in ITU-T history and demonstrates the global interest in discussing metaverse standardization.

In order to increase and spread interest in the metaverse, ITU is co-hosting forums with various standards development organizations, industry, and others, which is proving to be a great success, with more than 14,000 people attending a webcast session of the forum held in Shanghai, China in July 2023.

In addition, the FG-MV is organizing special sessions to promote the need for metaverse standardization and the major achievements of the FG-MV. The details for the events can be found on the FG-MV webpage,~\cite{ITU-T-FG-MV}.

Discussions on metaverse terminology and concept, CitiVerse, industry metaverse, domain-specific applications and services, infrastructure, platform interoperability, integration of real and virtual worlds, and security and privacy, etc., will be actively progressed at the ITU FG-MV, resulting in expecting the first major deliverables by the end of 2023.

\section{3GPP}
\label{sec:3GPP}

The 3GPP, or the 3rd Generation Partnership Project (3GPP), is responsible for the standardization of mobile telecommunication protocols. Its best-known work is the development and maintenance of the different cellular telephony standards such as: Global System for Mobile Communications (GSM, 2G), Universal Mobile Telecommunications Service (UMTS, 3G), Long Term Evolution (LTE-4G) and 5G NR (New Radio). The 3GPP organizes its working structures into three different streams, i.e., Radio Access Networks (RAN), Services and Systems Aspects (SA), and Core Network and Terminals (CT). Each of the three streams have multiple Working Groups, or WGs, focusing on standardizing different technical specifications. The 3GPP specifications cover cellular telecommunications technologies, including radio access, core network and service capabilities. They also provide hooks for non-radio access to the mobile core network, and for inter-working with non-3GPP networks. The 3GPP work spans normally 3 stages, namely the stage-1 requirements (in the SA1 WG), the stage-2 architecture and framework (in the SA2 - SA6, RAN WGs), and the stage-3 protocols (in the CT WGs). Roughly, each stage of a project will go thru both the study phase (Study Items are denoted SIDs) and the normative phase (memorialized in a Work Item Description, WID). 

\subsection{Metaverse in 3GPP}
\label{subsec:SA_Metaverse}

Metaverse has been used in various ways to refer to the broader implications of AR, VR or XR \& multi-modality services. Therefore, we will cover the Metaverse in a broader scope, including XR services. Currently in 3GPP, the Metaverse related SIDs and WIDs have spread to different Working Groups (WGs) belonging to different stages.
\ITUpar
In the stage-1 (SA1), there are three Metaverse related projects, i.e., the TACMM or TACtile and Multi-Modal communication service;  the high mobility for XR services; and the FS\_Metaverse or the localized mobile metaverse. Based on the stage-1 study conclusions and consolidated requirements, the stage-2 WGs (e.g., SA2 - SA6) have developed their own corresponding projects. The SA2 WG have two projects, namely the XRM or extended reality \& media services, and the XR \& Metaverse. The SA4 WG has multiple related projects, studying and exploring XR and AR use cases, Key Performance Indicators (KPIs), media formats, and more. \ITUpar
While both the SA2 and SA4 have 3GPP approved projects on XR \& Metaverse, the SA6 WG is still in the proposal phase for two potential projects, namely the network enabler for XR, and the FS\_MetaApp study. Comparably, the stage-3 (CT WGs) XR \& Metaverse work is fairly premature, for which only the CT3 WG has started the protocol studies.\ITUpar

\subsection{Metaverse in 3GPP SA1 - Stage 1}
\label{subsec:SA1_Metaverse}

The SA1 TACMM (of the 3GPP Release-18) is a predecessor project of the more-scoped Metaverse. It explores the scenario of a set of UEs for applications consisting of multiple types of flows from multiple type of devices, e.g. video/audio stream, and haptic or kinesthetic data for more immersive experience. Its QoS requirements bears high data rate and low latency KPIs. For example, the max allowed end-to-end latency is 5ms for both uplink and downlink between UE and the interface to data network, the data rate between 0.1 - 1.0 Gbps, and the reliability of 99.99\% in both UL and DL, etc~\cite{TS.22.261}. For the release-19, the new SID of supporting mobility for XR services studies the characteristics \& demands of mobile networks to support the XR services in high-speed mobile scenarios. 

Further, the real SA1 Rel-19 Metaverse project, or FS\_Metaverse, has completed the study on specific use cases and service requirements for localized mobile metaverse services to offer shared and interactive user experience of local contents and services, accessed either by users in the proximity or remotely. The AR services and content are associated with locations in the physical world. As the stage-1 work, FS\_Metaverse serves as the foundation for future studies of Metaverse services in other SA WGs, e.g., SA2, SA4, and SA6, etc. So far, it has finished studying the localized mobile metaverse services, avatar-based real-time communication, digital asset management, etc~\cite{TR.22.856}.

\subsection{Metaverse in 3GPP SA2 - Stage 2}
\label{subsec:SA2_Metaverse}

XRM or Extended Reality \& Media Services (of the 3GPP Rel-18) studies the key issues, solutions on how to support XR and advanced media services from a 3GPP architecture perspective. It explores and investigates various kinds of issues, e.g., network information exposure on congestion level information, data rate, delay difference and round-trip delay of QoS flow, estimated bandwidth, burst periodicity, along with the provisioning via 3rd-party provided and/or locally generated policies. This work has been concluded and normalized as of now~\cite{TS.23.501}.

The 3GPP Rel-19 will see the expansion of the XRM scope, i.e., a potentially new SID on XR \& Metaverse. It plans to support seamless XR streaming over 3GPP or non-3GPP devices. The proposal strives for solutions for multiplexed data flows as well as how to effectively process flow packets with the dynamic changes in traffic characteristics. There could be simultaneously multiple applications with multiple UEs in the system, with each UE having multiple flows on modalities, e.g., audio, video, haptic and kinesthetic. The pursued solutions are either highly likely related to existing IETF technologies (e.g., L4S) or require possible extensions to IETF technologies (e.g., differentiated handling of encrypted XRM streams).

\subsection{Metaverse in 3GPP SA4 - Stage 2}
\label{subsec:SA4_Metaverse}
The SA4 WG focuses more on the application layer interaction between the UE and video/audio media services~\cite{TS.26.501}. AR, VR along with media services being the foundations for the support and implementation of Metaverse, the 3GPP Rel-16 FS\_5GXR studied XR and AR device types, use case, KPIs, device architectures, media formats, call flows, and more. The rel-17 FS\_5GSTAR studied the end-to-end encoding, rendering, functional framework, transmission interaction, and KPI of AR/MR glasses. The rel-17 XRTraffic studied the traffic characteristics of different XR service types and how they are related to the 3GPP RAN-1 Rel-17 XR study item. The conclusions led to implications for the current rel-18 SA2 \& RAN XR projects. Further, multiple rel-18 SA4 projects explore the media capabilities of AR devices, the split rendering to enable UE to share rendering to edge, the optimization of the use of RTP for the uni-directional and bi-directional transport of real-time immersive media, and the support of a specific device type, tethering AR glass, etc.

\subsection{Metaverse in 3GPP SA6 - Stage 2}
\label{subsec:SA6_Metaverse}
The SA6 WG intends to study how the XR service can be enabled \& supported by network enabler layer based on the 3GPP system. The objectives revolve around fulfilling the high data rate and low latency requirement of XR services~\cite{TS.23.558}. Thus, a proposal has been raised recently, from the application-level (e.g. from the server point of view), to investigate and identify architecture requirements and solutions to enhance the service enabler application layer (SEAL) for better data delivery in order to support the XR services. A 2nd proposal, the FS\_MetaApp, plans to research the application enablement for localized mobile Metaverse services. Here, a multitude of new experiences, products and services, and other activities are expected to be enabled by the metaverse, also from the application enabler layer PoV. Some study items are specified, e.g., architecture support, functional model, etc. for application enablement under metaverse services in diversified areas like community, social, profession, and business, etc.

\subsection{Metaverse in 3GPP CT - Stage 3}
\label{subsec:CT_Metaverse}
So far, the XR \& Metaverse stage-3 work of CT is still at the very early phase since both the stage-1 and the stage-2 work are either on-going or just complete-in-phase now. They still need time to trickle down to the CT WGs. The XRM WID of the CT3 WG expanded the study of the potential impacts on NAS, or Non-access stratum. NAS is a functional layer in the NR, LTE, UMTS and GSM wireless telecom protocol stacks between the core network and user equipment. The CT3 also collaborates with the CT1 to help further support SA2 and RAN from the terminal side to continue the research on uplink data sending scenarios in the rel-18 XRM project.

\section{IETF}
\label{sec:IETF}

The Internet Engineering Task Force has not chartered any Metaverse Research or Working Group as of this writing. \ITUpar

In two IETF meeting, 115 in London in November 2022 and 117 in San Francisco in July 2023, side meetings were organized to present potential work items to the IETF community in some chartered Research or Working Group (RG or WG). \ITUpar

At this point, most of the related work to support the Metaverse is therefore distributed in different groups. \ITUpar

In the Information-Centric Networking Research Group, two drafts~\cite{aw-metaverse-icn-00-short,fmbk-icnrg-metaverse-01-short} have been presented to study the interactions of the Metaverse with Information-Centric concepts, such as the efficient distribution of information, or the separation of the content from a specific host server. Both consider the opportunities and challenges of implementing (and potentially integrating) the Metaverse with ICN tools and abstractions. 
Previous work~\cite{rfc9273,RFC7933} has looked at how ICN can support better multi-media~\cite{yu2015congestion,wang2014optimal,hu2019orchestrating,ramakrishnan2016adaptive,he2020efficient}  and immersive video-conferencing~\cite{zhang2017vr}.
\ITUpar

The Metaverse network demands are addressed in other existing working groups. For instance, bandwidth-hungry applications such as the Metaverse can leverage RFC8684~\cite{rfc8684-short} for multi-path TCP or~\cite{ietf-quic-multipath-05short} for multi-path QUIC connections to achieve higher or more reliable throughput. \ITUpar

Low-latency reliable media delivery solutions, such as Media-over-Quick~\cite{ietf-moq-requirements-01-short,jennings-moq-proto-00-short}, could also facilitate the deploymnent of Metaverse applications. \ITUpar

The APplication-aware Networking Working Group\footnote{https://datatracker.ietf.org/wg/apn/about/} attempts to provide application-dependent network services beyond best effort. \ITUpar

ALTO~\cite{rfc5693-short} similarly attempts to incorporate application-specific elements into the selection of a specific serving instance. 

The Computing-Aware Traffic Steering (CATS, https://datatracker.ietf.org/doc/charter-ietf-cats/) is also relevant inasmuch it attempts to incorporate server's information (compute metrics such as processing, storage
capabilities, and capacity) along with network information (bandwidth, latnecy) to steer traffic based on these metrics. Metaverse instances would benefit from such optimization.\ITUpar

However, APN, ALTO, CATS only consider the Metaverse as a generic application overlay and are not specifically optimized for it.

\section{MPAI}
\label{sec:MPAI}

The dozens of metaverse definitions that people have attempted over the years demonstrate that the word conveys different meanings to different people. We ourselves presented four in Section~\ref{sec:taxonomy}. Without trying to join the attempts, metaverse can be characterized as a system that captures and processes data from the real world, combines it with internally generated data, and creates virtual environments humans interact with. 

So far, metaverse designers have made independent technology choices, often without considering those of other developers. Recently, however, concerns have been raised about the metaverse as “walled gardens” because these do not fully exploit the opportunities conveyed by this word. The request is that metaverse should be “interoperable”. 

Since early 2022, MPAI – Moving Picture, Audio, and Data Coding by Artificial Intelligence  – the international, unaffiliated, non-profit organization developing standards for AI-based data coding, has pondered the issue of metaverse interoperability. It has done so because metaverse is a promising service area to which MPAI members can offer their expertise and because it can be seen as a platform allowing the integration of many data coding technologies, including some from MPAI.

Interoperability – the ability of a metaverse instance (M-Instance) to exchange and make use of the data of another M-Instance – needs its attributes to convey shared meanings. Direct interoperability refers to the case when the communication-enabling technologies are specified. This can hardly be applied to the current metaverse context because technologies are fast-evolving, and the number of ways they can be applied is just too large. Mediated interoperability refers to the case when data conversion services make up for incompatibilities created by the “everyone for himself” approach. This can work in simple cases, but conversion cannot cope with the large variety of technology combinations. Independently adopted data formats often cannot be converted to other independently adopted data formats.

MPAI has adopted the so-called functional interoperability. It has identified functionalities (some 150 of them) that an M-Instance exposes to another M-Instance or device in “Technical Report: MPAI Metaverse Model – Functionalities” (January 2023~\cite{MPAI-TR-F}). The large number of application domains has suggested the adaptation of the notion of Profile (a subset of the technologies used in a standard) to the metaverse case where technologies may come from several standards. These considerations are found in “Technical Report: MPAI Metaverse Model – Functionality Profiles”~\cite{MPAI-TR-FP}.

 “Technical Specification: MPAI Metaverse Model – Architecture”, published on 23 August 2023 with a request for Community Comments~\cite{MPAI-TR-A}, is a complete functional interoperability specification as applied to the metaverse. It contains:

 \begin{itemize}
 \item Scope: Specifies the coverage of the Technical Specification.
 \item Terms and definitions: A comprehensive collection of all terms used in the Specification (normative).
 \item Metaverse Functionalities: A revision of~\cite{MPAI-TR-F} documenting the functionalities supported by the Architecture Specification (informative).	
 \item Metaverse Operation Model: The Components of an M-Instance and the sequence of steps that are involved in Functionality provision. The Components are Processes performing Actions on Items (data and metadata supported by an M-Instance): Users (representing humans rendered as Personae), Devices (connecting humans with Users), Services, and Apps. Figure 1 depicts the relationships between Components (normative).
 \item Functional Requirements of Processes, Actions, Items, and Data Types collect the functional requirements of the four types of Process, the Actions that a Process can perform, the data and metadata, and the various data types used by an M-Instance (normative). 
 \item Use Cases verify the completeness and functionality of the Model in practical cases. Each of the nine use cases examines the applicability of Processes, Actions, Items, and Data Types (informative).
 \item Functionality Profiles identifies four profiles: Baseline supports metaverse activity without registration, Finance enables trading of assets, Management – a superset of Baseline and Finance – supports applications where Users manage rights to perform Actions, and High contains all other profiles with additional functionalities (normative).
\end{itemize}
 
Work is continuing with the development of a set of metaverse APIs in line with the Architecture Specification and the development of a “Table of Contents” for a future metaverse technology specification.

\startfigure
\includegraphics[width=\columnwidth]{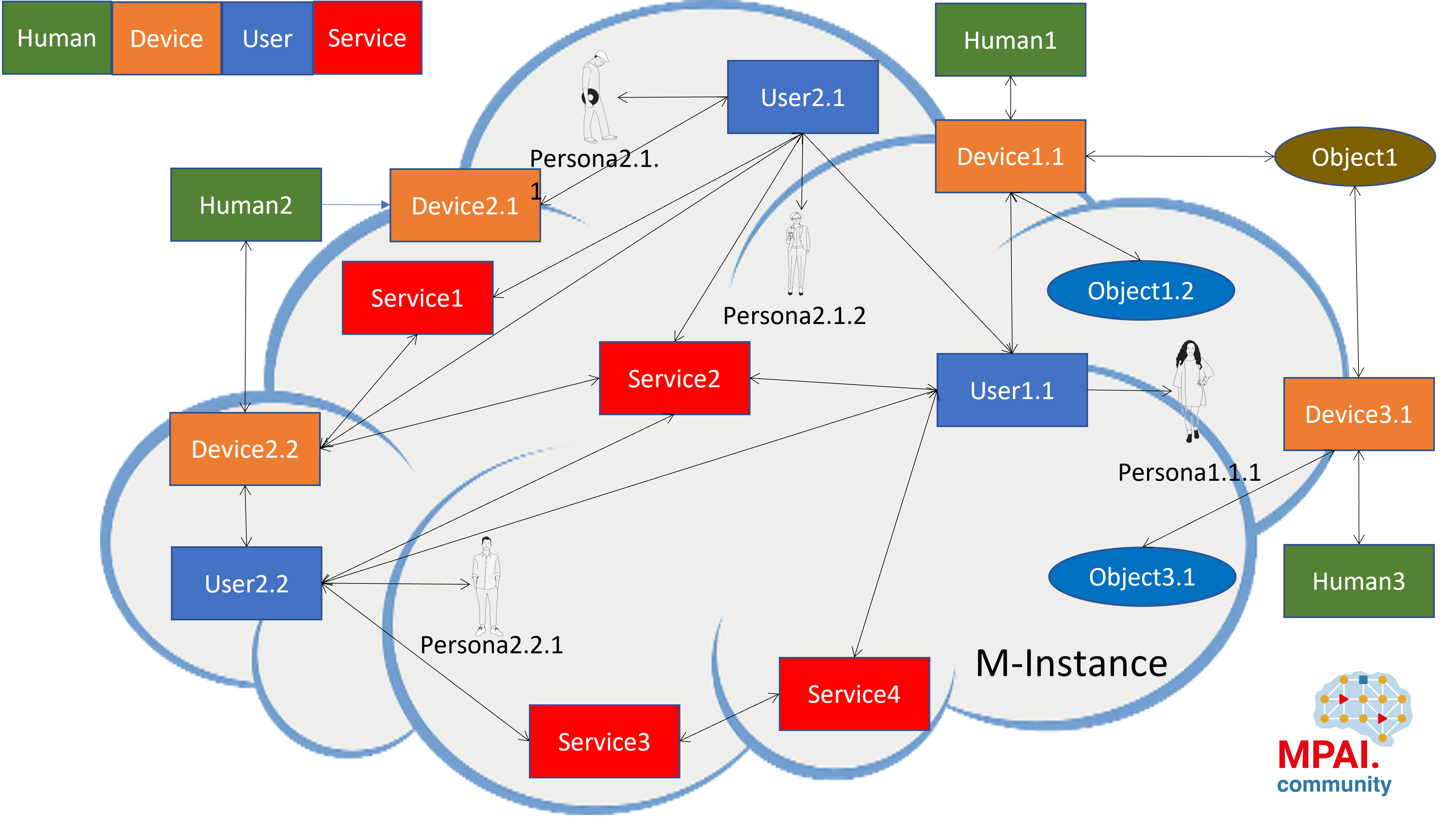}
\caption{A View of the MPAI Metaverse Model – Architecture}
\label{fig:MMM arch} 
\end{figure}

\section{IEEE and Others}
\label{sec:IEEE}

\subsection{IEEE}

In the IEEE, the IEEE Metaverse standards committee was created with the purpose of "{\it develop[ing] and maintain[ing] standards, recommended practices, and guides for metaverse, virtual reality and augmented reality, using an open and accredited process, and to advocate them on a global basis.}"

It is composed of two working groups, the IEEE Metaverse Working Group (CTS/MSC/MWG) and the IEEE Augmented Reality on Mobile Devices Working Group (ARMDWG). The Metaverse WG has so far came up with a Terminology, Definitions, and Taxonomy document~\cite{IEEE-P2048}.

In 2019, IEEE 2888 project was launched to standardize
interfaces between the cyber and physical worlds, which can be leveraged for Metaverse applications. IEEE 2888.1 moves sensory information from the physical world to the metaverse, and IEEE 2888.2 interfaces  from metaverse to physical space. IEEE 2888.3 standard can be used for the definition of digital things.

As in other SDOs, some standardization effort may find application to the Metaverse. As an example, IEEE 1589™ Standard for Augmented Reality Learning Experience Model, details an overarching integrated conceptual model for Augmented Reality, which is likely to be a key compenent of the metaverse.

\subsection{Others}
\label{sec:others}

{\bf OMI:} The Open Metaverse Interoperability Group (https://omigroup.org/) designs and promotes protocols for identity, social graphs, inventory, and more. These are at upper layer away from the infrastructure which is our focus. 

{\bf TIP:} The Telecom Infra Project (TIP) Metaverse Ready Networks is led by telcos to "{\it reimagine the way
we architect, build, deploy and manage networks}" in order to "{\it delivering metaverse-level
experiences at scale to wide audiences.}" (Quotes from~\cite{TIP}.)

The main goal of the project group is to converge towards ways of defining, providing and measuring a specifc level of QoE in the network and to 
use this to better support the Metaverse. 

{\em Metaverse Standards Forum:} This consortium (https://metaverse-standards.org/) intends to promote an open metaverse. It has set up several working groups, including ‘3D Asset Interoperability using USD and glTF’, ‘Digital Asset Management’, ‘Metaverse Standards Register’ and ‘Real/Virtual World Integration.’ It has more working groups under consideration, however none of them are related to the networking infrastructure. 

{\em Linux Foundation:} It has announced the creation of an Open Metaverse Foundation (https://www.openmv.org/), an "open, vendor-neutral community dedicated to creating open standards and software to support the open, global, scalable Metaverse." Within this foundation, the Networking Foundational Interest Group (FIG) focus on open source and standards pertaining to transport architecture, services, and operations.  

\section{Summary \& Comparison}
\label{sec:table}
\ITUpar

We attempt here to summarize the different efforts in Table~\ref{tab:summary} that compares all the different standards. 

The different SDOs work on different aspects, so there is no perfect alignment of the work across the SDOs. Rather, they have complementary efforts combined with inter-SDO collaboration. 

\begin{table*}
\centering
\caption{Comparison of the different Standardization Efforts}
\label{tab:summary}
\begin{tabular}{ |p{2.5cm}|p{5.5cm}|p{9cm}| } 
\hline
 Organization & Focus Areas & Description \\
\hline
\hline
\multirow{11}{8em}{ITU-T} & & \multirow{11}{24em}{The Focus Group on metaverse (FG-MV) was
established on 16 December 2022. The FG-MV
aims to lay the groundwork for international
standards that can help create an underlying
technology and business ecosystem. The
group analyses metaverse technical
requirements to identify key enabling
technologies in areas ranging from
multimedia and network optimization to
digital currencies, Internet of Things, digital
twins, and environmental sustainability. It
also provides a collaboration platform, for
identifying stakeholders with whom ITU-T
could collaborate, and for enabling the
inclusion of non-members to contribute to the
international technical pre-standardization
work.}\\  
&  - General Terminology \& definitions & \\
& - Applications \& Services & \\ 
& - Architecture \&
Infrastructure & \\
& - Virtual/Real World
Integration & \\
& - Interoperability & \\
& - Security, Data \& PII
Protection  & \\
& - Economic, regulatory \&
competition aspects  & \\
& - Sustainability, Accessibility \& Inclusion  & \\
&  - Collaboration with other
SDOs & \\
& & \\
\hline
\hline
\multirow{5}{8em}{3GPP} & - Stage-1 (SA1): Three
Metaverse related projects, i.e.,
the TACMM or tactile and multi-
modal communication service,
FS\_Metaverse or the localized
mobile metaverse and (WID)
Metaverse. & \multirow{5}{24em}{3GPP is responsible for the
standardization of mobile telecom
protocols. Its best-known work is the
development and maintenance of
GSM(2G), UMTS(3G), LTE(4G) and 5G NR.
The 3GPP work spans normally 3 stages,
namely the stage-1 requirements (in the
SA1 WG), the stage-2 architecture and
framework (in the SA2 - SA6, RAN WGs),
and the stage-3 protocols (in CT WGs).
Roughly, each stage of a project will go
thru both the study phase (SID) and the
normative phase (WID). Currently, the
Metaverse related SIDs and WIDs have
spread to different Working Groups
(WGs) belonging to different stages.} \\
& - Stage-2 (SA2 - SA6): & \\
& a) SA2 having two projects, the XRM or extended reality \& media services, and the XRM\_Ph2; & \\
& b) SA4 WG having multiple projects:
FS\_5GXR, FS\_5GSTAR,
XRTraffic, etc.; & \\
& c) SA6 having two projects,
FS\_AEXRS and FS\_MetaApp
study. & \\
\hline
\hline
\multirow{3}{8em}{IETF} & - Terms \& definitions & \multirow{3}{24em}{The IETF has not created a Metaverse Working Group yet,  Metaverse-related work is distributed over other WGs} \\
 & - Application Aware Networking & \\
 & - Infrastructure Support & \\
\hline
\hline
\multirow{15}{8em}{MPAI} & - Terms \& definitions & \multirow{15}{24em}{MPAI has started considering of standard opportunities in the metaverse area at the beginning of 2022. In the early months of 2023, MPAI published 2
Technical Reports of Functionalities and
Functional Profiles. In September it published Technical Specification: MPAI Metaverse Model (MPAI-MMM) – Architecture.
The specification enables Interoperability of two or more metaverse instances (M-Instances) {\em if} they rely on the Operation Model, and use the same Profile Architecture, {\em and} either the same technologies, {\em or} independent technologies while accessing Conversion Services that losslessly transform Data of an M-Instance A to Data of an M Instance B. All documents are available from (https://mpai.community/standards/mpai-mmm/)} \\ 
& & \\
& - Use Cases & \\
& & \\
& - Architecture & \\
& & \\
& - Operation Model & \\
& & \\
& - Functional Requirements & \\
& & \\
& - Functional Profiles & \\
& & \\ 
& - Reference Software & \\
& & \\ 
& - Collaboration with other SDOs & \\
\hline
\end{tabular}
\end{table*}

\section{Conclusion}
\label{sec:conclusion}
\ITUpar

We have provided a perspective on the Metaverse and its need from the network layer for its successful deployment. In particular, we proposed some definitions and a taxonomy to characterize the specific properties of a Metaverse instance, and to narrow down the standardization options. 

We then described current effort in standardization throughout multiple institutions, including ITU, 3GPP, MPAI, IETF, etc. We hope this overview will help the readers to navigate the standardization landscape. Providing common interfaces and protocols will be essential to the success of the Metaverse.


\printbibliography 

\ITUpar

\ITUpar

\end{document}